\documentclass[letterpaper, 10 pt, conference]{ieeeconf}

\IEEEoverridecommandlockouts
\usepackage{graphicx}
\usepackage{url}
\usepackage{todonotes}
\usepackage[per-mode = symbol]{siunitx}
\usepackage[english]{babel}
\usepackage{csquotes}
\usepackage[acronym]{glossaries}
\usepackage{amssymb}

\newacronym{EMG}{EMG}{electromyography}
\newacronym{sEMG}{sEMG}{surface EMG}
\newacronym{DOF}{DOF}{degree of freedom}
\newacronym{MVC}{MVC}{maximal voluntary contraction}
\newacronym{AUX}{AUX}{auxiliary}
\newacronym{NMF}{NMF}{non-negative matrix factorization} 
\newacronym{CNN}{CNN}{convolutional neural networks}
\newacronym{FFI}{FFI}{force feedback interface}
\newacronym{DoI}{DoI}{digit of interest}
\newacronym{GUI}{GUI}{graphical user interface}
\newacronym{CAD}{CAD}{computer-aided Design}

\title{\LARGE \bf
MyoKin3X: A Myoelectric Framework for\\Full-Hand 3D Force Recording
}

\author{Charlotte Rohleder$^{1}$, Raul C. S\^impetru$^{1}$, Annika W\"unsch$^{1}$, and Alessandro Del Vecchio$^{1}$%
\thanks{$^{1}$Neuromuscular Physiology and Neural Interfacing Laboratory, Friedrich-Alexander-Universit\"at Erlangen-N\"urnberg, 91052 Erlangen, Germany.}%
\thanks{This work was partially supported by the European Research Council (ERC) through the project GRASPAGAIN under grant 101118089, 
by the German Federal Ministry of Education and Research (BMBF) through the project MYOREHAB under grant 01DN23002, by the German Federal Ministry of Research, Technology and Space (BMFTR) through the project BIONIK under grant 16SV9639, 
and through the Deutsche Forschungsgemeinschaft (DFG) under grant 523352235 to ADV.}%
}

\begin{document}

\flushbottom
\maketitle
\thispagestyle{empty}

\begin{abstract}
Simultaneous multi-directional force measurement across all five digits is essential for studying hand coordination, compensatory forces, and myoelectric control, yet existing systems typically trade off digit coverage, force dimensionality, and anatomical adaptability.
Reliable full-hand acquisition remains challenging because multi-axis calibration, hand-size adjustment, and consistent digit-specific force reconstruction are technically demanding.
We present MyoKin3X, a customizable full-hand framework for simultaneous 3D force measurement of up to five digits that provides robust and validated force reconstruction.
It combines an anatomically versatile structure with five integrated 3D force sensors and a standalone software for synchronized electromyography (intramuscular and surface) and force acquisition.
MyoKin3X provides in-place cross-calibration of all five force sensors, single- and multi-digit maximal voluntary contraction recording, and automated coordinate transformation to digit-specific coordinate systems for standardized analysis across subjects and tasks.
Calibration validation demonstrates high stability of the axis-specific calibration factors, 
with a mean coefficient of variation of $0.04\%$.
The maximum force error was $\pm 0.06$\,N at 50\,N. 
It also shows effective inter-axis decoupling (crosstalk reduction: $92.71\%$ mean reduction; residual crosstalk below $0.02\%$ for most axis pairs) and high predictive accuracy ($R^2 \geq 0.99$ across all sensors).
The software includes four feedback modes: 1D ramps, fatigue protocols, 2D arbitrary target ramps, and 2D exploratory tasks.
MyoKin3X therefore enables standardized full-hand force acquisition with validated measurement reliability, flexible protocol control, and real-time visualization for high-fidelity studies of hand motor control, muscle synergies, and rehabilitation-oriented human-machine interfacing.
\end{abstract}

\section{Introduction}

The human hand is actuated by more than 30 muscles that enable dexterous force generation across multiple coupled \glspl{DOF}, making the study of coordinated force production a central challenge in neurophysiology and biomechanics~\cite{Malesevic2019InstrumentedContractions}. 
\Gls{EMG} provides direct access to muscle and motor-neuron activity~\cite{Zajac1992, Hepp-Reymond1996PrecisionHumans, DelVecchio2020}, but linking these neural commands to their mechanical output requires precise and simultaneous force measurements at the digit level.

Numerous force measurement systems have been developed to study multi-digit force production, including load-cell-based systems for flexion-extension measurements~\cite{Reilly2003IncompleteStudy, Grison2024IntramuscularControl}, multi-axis transducers for natural-posture testing~\cite{Pataky2007AHand}, and dynamometers for isolated finger segments~\cite{Irwin2013DevelopmentForce, Castellini2014AForces, Ross2023DesignMeasurement}.
However, these systems typically trade off digit coverage, force directions, and anatomical adaptability~\cite{Malesevic2019InstrumentedContractions}, because hand-size variability, multi-axis calibration, and accurate force reconstruction is technically demanding.

Kapur et al.~\cite{Kapur2010FingerTask} showed that digits produce substantial off-axis forces in single- and multi-digit tasks, and that multi-dimensional feedback enhances control by preserving task-relevant information. 
These requirements extend to rehabilitation-oriented studies of hand motor dimensions~\cite{simpetru_identification_2024} and human-machine interfacing applications requiring simultaneous proportional multi-digit and multi-\gls{DOF} control~\cite{simpetru_proportional_2023}.

\begin{figure*}[!t]
    \centering
    \includegraphics[width=\textwidth]{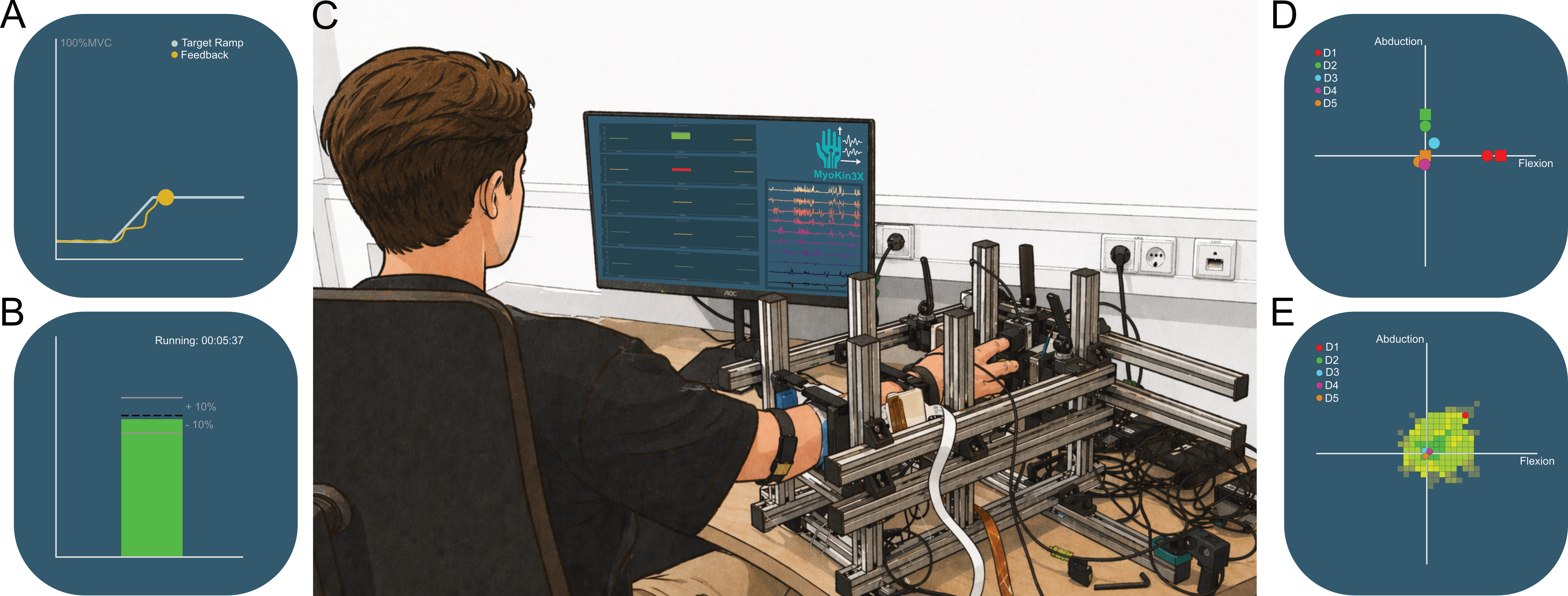}  %
    \caption{\textbf{Framework Overview.}  
    \textbf{A)} 1D ramp feedback showing target and exerted task-related force magnitude (digit component breakdown shown in main setup sub-figure for clarity). 
    \textbf{B)} 1D fatigue feedback showing exerted magnitude as a bar, target as a black dotted line, and acceptable deviation range as two grey lines. 
    \textbf{C)} Drawing of the MyoKin3X setup with the 3D force framework and the visual interface displaying electromyographic signals and force distribution per sensor axes.
    \textbf{D)} 2D feedback in each digit's flexion-abduction plane; colored dots show exerted force per digit and colored squares show predefined targets. 
    \textbf{E)} Exploratory 2D flexion-abduction feedback with digit dots and a heatmap of visit counts per bin.}
    \label{fig:overview}
\end{figure*}

To address this gap, we introduce MyoKin3X, a versatile and customizable full-hand 3D force-recording framework for simultaneous multi-digit force acquisition with synchronized \gls{EMG} (Fig.~\ref{fig:overview}).
It combines five 3-axis force sensors in an anatomically adjustable structure with standalone software for acquisition and analysis.
Key features include single- and multi-digit \gls{MVC} measurement, built-in calibration, automated digit-specific coordinate transformation, and four feedback paradigms.
Together, these components provide a standardized methodology for multi-digit force acquisition in motor-control, force synergy, and myoelectric-interfacing studies.

\section{Methods and Materials}

\subsection{Specification of the Force Sensing Setup}

Digit forces are recorded using five custom-mounted three-axis load cells (K3D40 \SI{50}{\newton}, ME-Me{\ss}systeme GmbH, Hennigsdorf, Germany) with compact dimensions (40 × 40 × 20 \si{\milli\metre}) to facilitate integration into the experimental setup. 

Each sensor provides a nominal range of $\pm 50$\,N per axis and is connected to a multi-channel strain-gauge amplifier (GSV-1A4, ME-Me{\ss}systeme GmbH), configured to an input sensitivity of $4$\,mV/V and supplying a bridge excitation voltage of $5$\,V. 
Under this configuration, the sensor sensitivity is approximately $18.5$\,N/V per axis.

The conditioned analog signals are acquired through the \gls{AUX} inputs of a bioamplifier system (OT Bioelettronica, Torino, Italy), which provide a $\pm 5$\,V input range and 16-bit analog-to-digital conversion. 
This corresponds to a theoretical digital force resolution of approximately $2.8$\,mN per least significant bit. 
Sampling is performed synchronously with the EMG acquisition at $\SIrange{2}{10}{\kilo\hertz}$.
Considering the specified sensor and amplifier tolerances, the practical system accuracy is estimated to be on the order of $\pm 0.25$ to $\pm 0.30$\,N per axis under static loading conditions.

The setup can accommodate other sensor models if the output range matches the recording device.
The K3D40 sensor was chosen for its high accuracy and sensitivity while maintaining compact dimensions.

\subsection{Configuration of the 3D Force Framework}

\begin{figure*}[!t]
    \includegraphics[width=\textwidth]{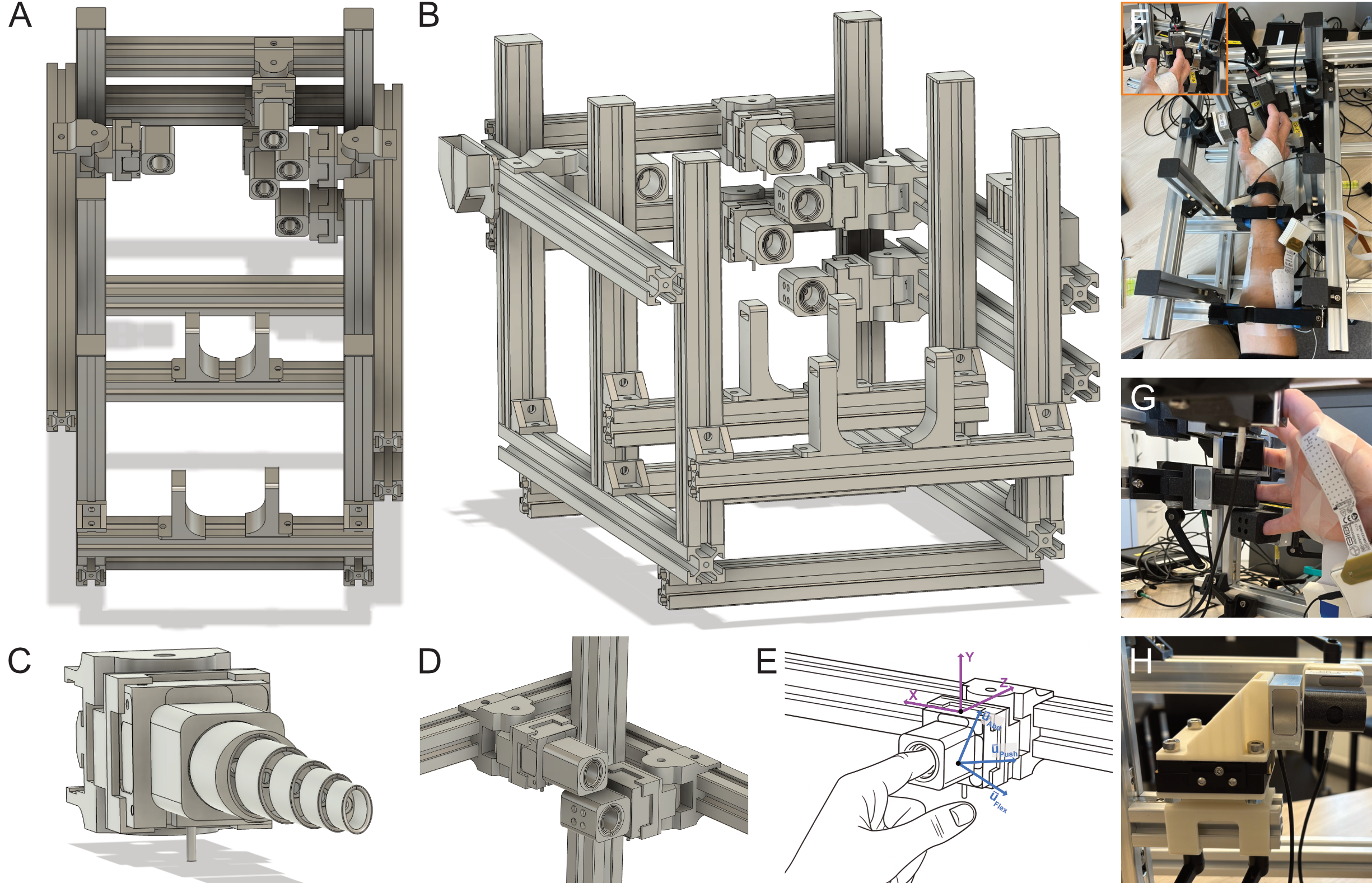}  %
    \caption{\textbf{Force Framework Setup Overview and Components. A)} Top computer-aided design (\gls{CAD}) view with five mounted sensors and two arm-support profiles. 
    \textbf{B)} Front-left \gls{CAD} view of the force framework with force sensor and intramuscular \gls{EMG} amplifier holder. 
    \textbf{C)} \gls{CAD} view of one sensor assembly with digit box and telescopic inlets of different size. 
    \textbf{D)} Close-up \gls{CAD} view with two digit boxes mounted on different sides of two sensors; each sensor rotates horizontally via the central fixation joint (quick-release levers not shown). 
    \textbf{E)} Schematic of transformation from the sensor coordinate system (purple, manufacturer axes) to the digit-specific system (blue: flexion, abduction, push) aligned with digit's degrees of freedom. 
    \textbf{F)} Photograph of a right-handed participant in the framework with all digits inserted and arm fixation at wrist and elbow; surface \gls{EMG} grids cover flexor, extensor, and intrinsic hand muscles. In the top left, close-up of inserted digits.
    \textbf{G)} Close-up palmar view of the hand with all five digits inserted in digit boxes. 
    \textbf{H)} In-Place sensor calibration setup with the reference sensor mounted on a linear stage to apply controlled loading along the x-axis.}
    \label{fig:frameworkParts}
\end{figure*}

The framework is adjusted to the participant's hand dimensions and dominance (Fig.~\ref{fig:frameworkParts}). 
It is built from aluminum profiles (Nut 8; cross-section: $\SI{30}{\milli\metre} \times \SI{30}{\milli\metre}$; lengths: \SI{30}{\centi\metre} and \SI{50}{\centi\metre}) with custom 3D-printed components and quick-release connections for fast reconfiguration.

Profiles, sensors, and wrist/elbow supports can be aligned to the participant's anatomy, and measuring tapes on the profiles support reproducible setup documentation.
To ensure a precise fit up to the distal interphalangeal joint, the appropriate inlet size for each finger is determined by trial fitting.
The base inlet can be rotated inside the digit box to facilitate a comfortable insertion angle.

The framework is mounted on a height-adjustable table so participants can sit upright with the elbow supported at \SI{90}{\degree}.
For intramuscular \gls{EMG} recordings, A/D adapters can be affixed to the proximal profiles, and force sensor connectors can be secured in custom-made holders for transport.

\subsection{Force Sensor Calibration System Design and Workflow}\label{sub:force_cali_software}

The calibration system is integrated into the force-recording framework, enabling in-place calibration of all five three-axis sensors under their final mounting conditions. 

Calibration is performed using a modular aluminum fixture with a precision linear stage and an external three-axis reference sensor (K3D40 \SI{50}{\newton}). 
By repositioning the stage relative to the sensor's digit box, controlled loading is applied along the target axis (Fig.~\ref{fig:frameworkParts}H).

For each sensor and axis, three loading-unloading cycles are recorded with synchronous acquisition of reference and test signals. 
Channel offsets are removed prior to each trial, and voltages are converted to force using manufacturer calibration factors.
Axis-wise calibration factors are obtained as the slope of a linear regression between reference force and test-axis voltage, with repeatability quantified by the coefficient of variation across repetitions.

To account for inter-axis coupling, a sensor-specific \(3 \times 3\) decoupling matrix is estimated via multivariate linear regression using all sensor outputs. 
The resulting matrix is applied sample-wise to transform raw voltages into corrected three-dimensional force estimates in Newton. 
Calibration performance is evaluated based on the coefficient of variation of calibration factors, crosstalk reduction, and predictive accuracy ($R^2$) on independent test data.

\subsection{Data Processing and Graphical User Interface}

Data acquisition uses a bioamplifier (OT Bioelettronica, Torino, Italy) and custom PySide6 application streamed through OTB Light and the Biosignal Device Interface~\cite{BiosigDeviceInterface}.

The \gls{GUI} records \gls{EMG} and \gls{AUX} signals, loads protocol settings from JSON, and provides tabs for connection, visualization, calibration, \gls{MVC} recording, and force feedback. 
It supports optional real-time \gls{EMG} visualization, in-session force sensor calibration, offset recording, and live plotting of all fifteen \gls{AUX} channels. %

\gls{MVC} values for each digit's \gls{DOF} directions, as well as for predefined multi-digit contractions, can be recorded to subsequently calculate digit-specific coordinate transformations (Section~\ref{sub:digit_transform}).
Each \gls{MVC} recording lasts \SI{5}{\second}. 
If no \gls{MVC} recording is performed, the system defaults to the sensor’s native coordinate system for all subsequent recordings.

Once \gls{MVC} values and coordinate transforms are computed, one of four configurable \glspl{FFI} is selected, each supporting different experimental objectives and varying in visualization detail and flexibility (Sections~\ref{ffi_circle} to~\ref{ffi_heatmap}).
All recordings are stored as zarr arrays containing \gls{EMG}, \gls{AUX}, and, where applicable, predefined force-feedback signals.

\subsection{Digit-specific Coordinate Transform}\label{sub:digit_transform}
For each subject, we estimate a digit-specific coordinate system spanned by flexion, abduction, and push directions (Fig.~\ref{fig:frameworkParts}E).
Raw \gls{AUX} samples are first converted from device units to Newton using the previously estimated decoupling matrices, establishing a calibrated 15D force measurement space in Newton. 
For each digit, only its corresponding sensor's three channels are used to estimate a local basis. 
This yields a unitless projection tensor
\begin{equation}
\mathbf{W}\in\mathbb{R}^{5\times 3\times 15},
\end{equation}
which maps the 15D \gls{AUX} space (in Newton) into a standardized 3D force representation for each digit $d\in\{1,\dots,5\}$. 
To estimate this projection, the \gls{MVC} repetitions for each contraction type are loaded and smoothed (centered moving average, \SI{50}{\milli\second}). Let $\mathbf{f}_r(t)\in\mathbb{R}^{3}$ denote the local force vector at sample index $t$ in repetition $r$. 
With threshold parameter $\eta=0.85$, we define the corresponding high-force sample index set
\begin{equation}
\mathcal{P}_r = \{t\;|\;\|\mathbf{f}_r(t)\|_2 \ge \eta\,\max_t\|\mathbf{f}_r(t)\|_2\},
\end{equation}
and compute the mean unit direction for each repetition as
\begin{equation}
\hat{\mathbf{u}}_r = \operatorname{norm}\!\left(\sum_{t\in\mathcal{P}_r} \frac{\mathbf{f}_r(t)}{\|\mathbf{f}_r(t)\|_2}\right).
\end{equation}
The final contraction direction is then obtained by averaging these repetition-wise directions and normalizing:
\begin{equation}
\mathbf{u}=\operatorname{norm}\!\left(\frac{1}{R}\sum_{r=1}^{R}\hat{\mathbf{u}}_r\right).
\end{equation}

If coordinate transformation is enabled in the \gls{MVC} tab, flexion and abduction are estimated from their dedicated recordings. To enforce independent axes, the component of the raw abduction direction along flexion is removed and the result is normalized:
\begin{equation}
\mathbf{u}_{d,\mathrm{Abd}} = \operatorname{norm}\!\left(\mathbf{u}_{d,\mathrm{Abd}}^{\ast} - \operatorname{proj}_{\mathbf{u}_{d,\mathrm{Flex}}}\!\left(\mathbf{u}_{d,\mathrm{Abd}}^{\ast}\right)\right),
\end{equation}
and push is defined by
\begin{equation}
\mathbf{u}_{d,\mathrm{Push}} = \operatorname{norm}\!\left(\mathbf{u}_{d,\mathrm{Flex}} \times \mathbf{u}_{d,\mathrm{Abd}}\right).
\end{equation}
If coordinate transformation is disabled, flexion is the dominant signed sensor axis, and abduction/push are assigned to the remaining non-overlapping axes.

\gls{MVC} magnitude $m$ is always computed by projection:
\begin{equation}
m_r = \max_t\,\big|\mathbf{u}^\top\mathbf{f}_r(t)\big|.
\end{equation}
Final \gls{MVC} values are aggregated across repetitions (mean by default, optionally max or median). 
Flexion, abduction, and push \gls{MVC}s use their basis vectors; extension, adduction, and pull are computed by projection onto the same respective axes.
The \glspl{MVC} of digits for which the flexion \gls{MVC} is not recorded are set to \SI{1}{\newton} by convention.

For multi-digit contractions (2-pinching, 3-pinching, fist), the same projection tensor $\mathbf{W}$ is used to project into the digit-specific coordinate system. 
For task $q\in\{1,\dots,N_{\mathrm{MDC}}\}$ and repetition $r$, the projected force is
\begin{equation}
\mathbf{F}_{q,r}(t)\in\mathbb{R}^{5\times 3}.
\end{equation}
To isolate the active digits of this task, a masked tensor $\widetilde{\mathbf{F}}_{q,r}(t)$ is formed by keeping only entries for $d\in\mathcal{D}_q$ and setting all other digit entries to zero, where $\mathcal{D}_q$ is the active-digit set of task $q$. 
The global activity signal is then
\begin{equation}
g_{q,r}(t)=\|\operatorname{vec}(\widetilde{\mathbf{F}}_{q,r}(t))\|_2,
\end{equation}
and the corresponding high-force sample index set is
\begin{equation}
\mathcal{P}_{q,r}=\{t\;|\; g_{q,r}(t)\ge\eta\,\max_t g_{q,r}(t)\}.
\end{equation}
For component index $k\in\{\mathrm{Flex},\mathrm{Abd},\mathrm{Push}\}$, the per-repetition multi-digit \gls{MVC} is
\begin{equation}
m_{q,r,d,k}=\max_{t\in\mathcal{P}_{q,r}}\left|\widetilde{F}_{q,r,d,k}(t)\right|.
\end{equation}
Finally, $m_{q,r,d,k}$ is aggregated across repetitions $r$ (mean by default, optionally max or median), yielding a multi-digit \gls{MVC} tensor of shape $N_{\mathrm{MDC}}\times 5\times 3$.

\subsection{Force Feedback Computation}\label{sub:compute_force_feedback}
All \glspl{FFI} share the same online preprocessing. Let $\mathbf{a}(t)\in\mathbb{R}^{15}$ denote the raw \gls{AUX} vector. 
Data are decoupled, converted to Newton, and projected with the previously estimated tensor $\mathbf{W}$ (Section~\ref{sub:digit_transform}):
\begin{equation}
\mathbf{F}(t)=\mathbf{W}\,\mathbf{a}(t).
\end{equation}
For display, force is averaged over a chunk of $N$ samples,
\begin{equation}
\bar{\mathbf{F}}=\frac{1}{N}\sum_{t=1}^{N}\mathbf{F}(t).
\end{equation}
The full transformed 3D force vector is retained for recording and downstream analysis in all interfaces; only the real-time visualization differs between feedback modes.

The 2D interfaces (Sections~\ref{ffi_vector} and~\ref{ffi_heatmap}) visualize the projection of $\bar{\mathbf{F}}$ onto each digit's Flex/Abd plane while preserving the push component in the stored transformed signal. 
Targets are provided per digit and axis from predefined JSON trajectories, and feedback can be displayed either in absolute units (Newton) or normalized units (\%MVC).

The 1D ramp interface (Section~\ref{ffi_circle}) computes a scalar force value from the task-relevant components of the selected digits. For each participating digit $d$, let $\mathcal{K}_d$ denote the selected set of local directions (e.g., flexion alone, or flexion and abduction for a 2D target). The per-digit task magnitude is
\begin{equation}
f_d=\left\|\bar{\mathbf{F}}_{d,\mathcal{K}_d}\right\|_2,
\end{equation}
and the task-specific cumulative force is
\begin{equation}
F_q=\sum_{d\in\mathcal{D}_q} f_d.
\end{equation}
Let $m_{q,d,k}$ denote the selected task $q$'s \gls{MVC} for digit $d$ and local direction $k$, with inactive components set to zero. For a selected component set $\mathcal{K}_d$, we write $m_{q,d,\mathcal{K}_d}$ for the vector of corresponding component-wise \glspl{MVC}. The task-specific cumulative MVC-value is
\begin{equation}
F_{q,\mathrm{MVC}}=\sum_{d\in\mathcal{D}_q}\left\|m_{q,d,\mathcal{K}_d}\right\|_2,
\end{equation}
yielding normalized ramp feedback
\begin{equation}
F_{q,\%\mathrm{MVC}}=100\,\frac{F_q}{F_{q,\mathrm{MVC}}}.
\end{equation}
In contrast, the fatigue interface (\ref{ffi_fatigue}) monitors one selected digit-direction pair $(d^\ast,k^\ast)$ directly rather than the multi-axis task magnitude:
\begin{equation}
F_{\mathrm{FTG}}=\bar{F}_{d^\ast,k^\ast},
\qquad
F_{\mathrm{FTG},\%\mathrm{MVC}}=100\,\frac{F_{\mathrm{FTG}}}{m_{q,d^\ast,k^\ast}}.
\end{equation}
The displayed value is treated as a non-negative \%MVC magnitude and compared online against a target band.
Recording is stopped automatically if deviation exceeds the tolerance for a configurable duration.

For multi-digit tasks, the same computations are retained to enable a common processing path across feedback paradigms and allow direct comparison across interaces.

\section{Results}

\subsection{Design of a 3D Multi-Digit Force Recording Framework}

The framework comprises two main side panels connected by central struts. 
In the right-handed configuration, as shown in Fig.~\ref{fig:frameworkParts}, the thumb sensor is mounted on the left panel, while the remaining sensors alternate between rear and right struts to maintain spacing. 
Each sensor assembly includes a 3-axis sensor, a stator profile mount, and a digit box attached to the measuring unit. 
The mount has three parts: rail integration, horizontal alignment via two adjustable components, and connection to the rotating element. 
Sensor angle is fixed with a quick-release clamp that compresses the central components through a threaded nut. 
Digit boxes consist of a sensor-mounted main housing and interchangeable elliptical inlets. 
Depending on sensor position, the housing mounts on the sensor rear or side. 
Nested inlets adapt box size to finger dimensions and are fixed by a rear screw. 
Two front horizontal bars provide wrist and elbow fixation; padded Velcro-adjustable sliders stabilize the arm during measurements.

\subsection{Calibration Validation and Sensor Decoupling}
\label{sub:calibration_results}
Across repeated calibrations in the same mounting configuration, calibration factors for all five sensors and three axes showed a mean coefficient of variation of $0.0387\%$. 
At practical force levels, this corresponds to maximum errors of $\pm 0.063$\,N at \SI{50}{\newton} and $\pm 0.013$\,N at \SI{10}{\newton}, relative to a system resolution of approximately \SI{2.8}{\milli\newton}. 
Before decoupling, raw sensor outputs showed inter-axis coupling of $0.381\%$ to $3.602\%$ on the off-diagonal elements. 
Applying the sensor-specific \(3 \times 3\) decoupling matrix reduced inter-axis coupling by a mean of $92.71\%$ across all sensors and axis pairs, with residual crosstalk below $0.02\%$ for most axis pairs and a maximum residual of $0.175\%$ on a single axis pair. 
Validation on an independent test set yielded coefficients of determination of $R^2 \geq 0.9997$ across all sensor axes.

\begin{figure}[!t]
    \centering
    \includegraphics[width=0.99\columnwidth]{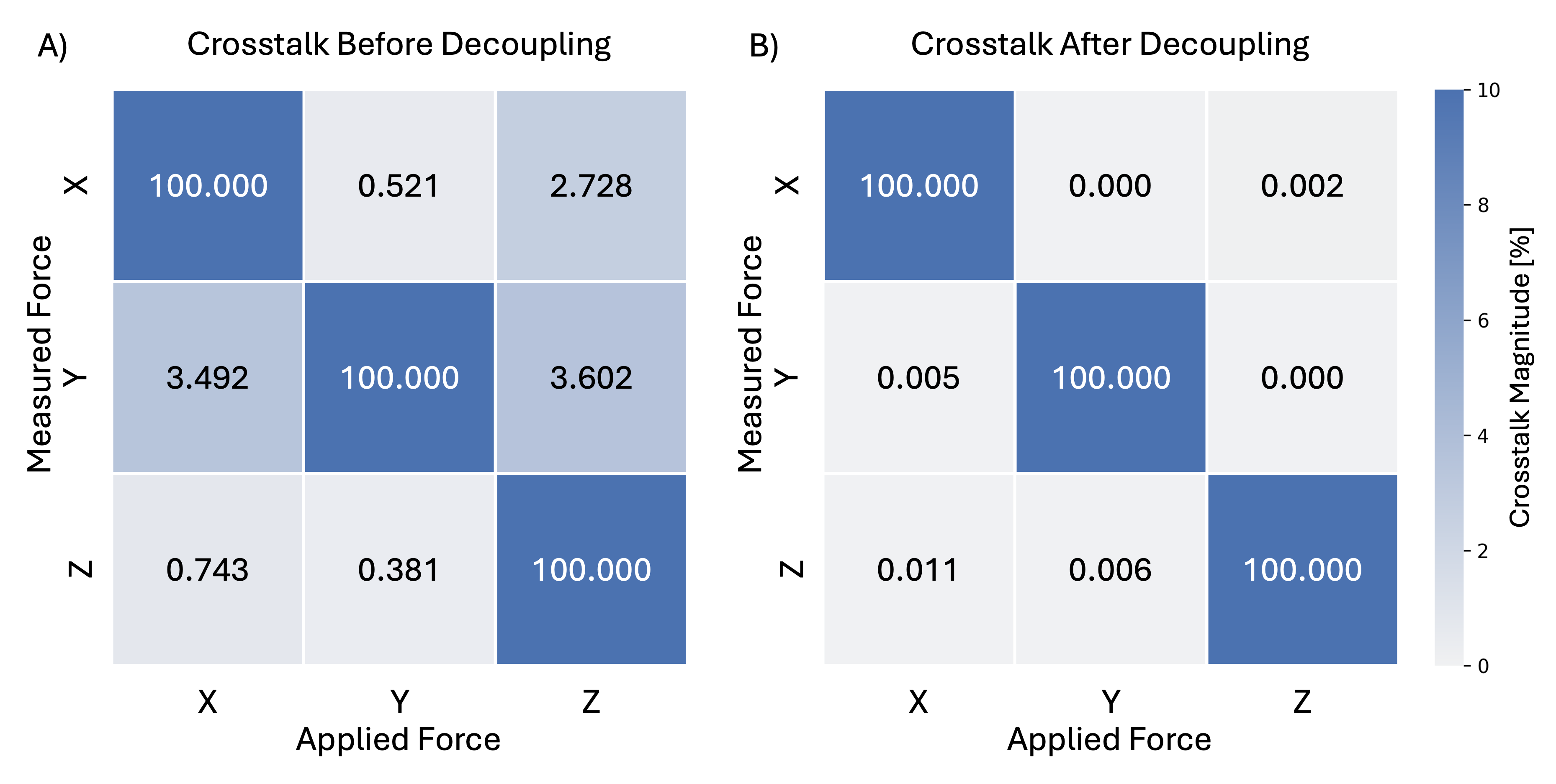}
    \caption{\textbf{Sensor Decoupling Effectiveness.}
    Crosstalk analysis for one sensor showing inter-axis coupling before and after decoupling. 
    \textbf{A)} Before decoupling, inter-axis coupling ranges from $0.381\%$ to $3.602\%$ on off-diagonal elements. 
    \textbf{B)} After decoupling, residual crosstalk is reduced to $0\%$--$0.011\%$, while the worst-case residual across all sensors is $0.175\%$. 
    }
    \label{fig:crosstalk}
\end{figure}

\subsection{The Software Interface}

The software interface provides a unified workflow for all four \glspl{FFI} with shared signal processing and task-specific visualization. 
In all modes, streamed \gls{AUX} data are converted, decoupled, transformed into the digit-specific coordinate system, and displayed as absolute force (N) or normalized units (\%\gls{MVC}) (Section~\ref{sub:compute_force_feedback}). 
Users configure task parameters from JSON-defined protocols, select active digits and \glspl{DOF} of interest, receive real-time feedback, and record all \gls{AUX} and \gls{EMG} channels for later analysis.

Across interfaces, the application supports manual start/stop, automatic termination by protocol rules (e.g., duration or tolerance violations), and standardized export of synchronized \gls{EMG}, \gls{AUX}, and task-related signals to timestamped zarr files. 
For \gls{FFI} 3 and 4, the user can flip the x-axis (flexion/extension) using a checkbox to align cursor motion with the perceived natural pushing direction.
The four interfaces differ primarily in the feedback objective and display paradigm: \gls{FFI} 1 compresses task-relevant force to a scalar ramp with per-digit decomposition, \gls{FFI} 2 monitors sustained single-direction performance for fatigue studies, \gls{FFI} 3 provides target-guided control in digit-specific 2D force space, and \gls{FFI} 4 enables free 2D exploration with cumulative heatmap coverage.

\subsubsection{FFI 1 -- 1D Force Ramp}\label{ffi_circle}

The first \gls{FFI} converts multi-dimensional digit forces into a 1D feedback signal (Fig.~\ref{fig:overview}A). 
The task force $F_q$ (Section~\ref{sub:compute_force_feedback}) is the magnitude of one selected digit or, for multi-digit tasks, the sum across active digits. 
The interface shows a rolling 2-second target preview, the current force point, and the previous 2 seconds for comparison. 
A second panel shows force distribution in five subplots with three bars each (flexion, abduction, push), including positive/negative directions, per-direction target markers, and color-coded error thresholds (green: $<5\%$, yellow: $<10\%$, red: $\geq10\%$). 
Tasks are loaded from JSON trajectories (ramp or arbitrary) and only selected digits and \glspl{DOF} contribute to $F_q$. 
The ramp is shown in \%\gls{MVC}, while component bars remain in Newton.

\subsubsection{FFI 2 -- Fatigue}\label{ffi_fatigue}
The second interface targets fatigue studies and displays one selected digit-\gls{DOF} (Fig.~\ref{fig:overview} B). 
The \gls{DOF}'s magnitude is shown as a bar with a black dotted target line and dotted tolerance bounds (default $\pm 10\%$ of target); the bar is green within bounds and red outside. 
Recording stops automatically if deviation exceeds tolerance for longer than a configurable duration (default \SI{3}{\second}). 

\subsubsection{FFI 3 -- 2D Vector Space}\label{ffi_vector}
The third interface extends feedback to all five digits in each digit's flexion-abduction plane (Fig.~\ref{fig:overview} D). 
All digits remain visible during the experiment, each represented by a colored circle positioned according to its force in the selected \glspl{DOF}. 
Per-digit targets are shown as colored squares, and participants track one or more targets by adjusting force in the relevant digit's force space. 
Arbitrary 2D trajectories are configurable per digit through holding times, target points in \%\gls{MVC}, and travel speed in \%\gls{MVC}/second (Section~\ref{sub:compute_force_feedback}).
Feedback can be displayed in Newton or normalized by \gls{MVC} units, enabling target-guided single- and multi-digit coordination in a shared 2D space.

\subsubsection{FFI 4 -- 2D Heatmap Exploration}\label{ffi_heatmap}
The fourth \gls{FFI} is an exploratory mode for sampling the full contraction space of the hand. 
Using the 2D coordinate system from Section~\ref{ffi_vector}, it removes predefined targets and asks participants to freely explore for a set duration (default \SI{20}{\second}) while visualizing the flexion-abduction projection of the digit-specific 3D force signal. 
A pixel-based heatmap is generated per predefined single- and multi-digit contraction tasks and updated across recordings (Fig.~\ref{fig:overview}E), quantifying visit frequency via color-coding. 
Only the digits of interest affect the heatmap, while all five digits remain visible for validation and analysis.

\section{Discussion}

MyoKin3X addresses a key gap in multi-digit kinetics research by enabling simultaneous five-digit 3D force acquisition with synchronized \gls{EMG} in a single anatomically adjustable platform. 
Prior systems provided important advances in digit-level measurement and configurability but required trade-offs in digit count, dimensionality, or flexibility~\cite{Pataky2007AHand, Malesevic2019InstrumentedContractions, Grison2024IntramuscularControl}. 
By preserving full multi-directional force information, MyoKin3X supports the study of off-axis force production, compensation, and inter-digit coordination~\cite{Kapur2010FingerTask}. 
In-place cross-calibration and sensor-specific decoupling further enable reproducible force reconstruction across sessions.

Calibration results demonstrate high stability (mean coefficient of variation $0.0387\%$) and effective crosstalk suppression (mean reduction $92.71\%$, worst-case residual $0.175\%$). 
Independent validation shows high predictive accuracy ($R^2 \geq 0.9997$), indicating robust 3D force estimation. 
Under controlled conditions, these results are consistent with prior multi-axis sensing studies and support the use of explicit decoupling approaches~\cite{li_multiaxis_2025, kebede_decoupled_2019, al-mai_design_2018}. 
However, direct comparison with patient-oriented hand-force studies remains limited and should be addressed in future validation efforts~\cite{nestler_reliability_2019}.

Beyond hardware, the integrated software framework combines calibration, coordinate transformation, protocol control, and real-time feedback. 
Four interfaces support both conventional and exploratory protocols while storing the full myoelectric and 3D force data per digit. 
Projection into digit-aligned coordinate systems and standardized data acquisition aim to improve cross-subject comparability and facilitate downstream motor-control and machine-learning analyses.

Limitations include unquantified long-term calibration stability, potential residual proximal-motion contributions despite stabilization, and a sensor range that prioritizes sensitivity over high-force capacity. 
Future work should evaluate longitudinal stability, incorporate proximal sensing for improved decoupling, and, given the modular hardware and sensor-agnostic software, assess higher-range sensors that maintain low-force sensitivity. 
Integrating online \gls{EMG}-based force estimation will further enable evaluation of myoelectric control in realistic tasks.

\section{Conclusion}

MyoKin3X introduces a unified framework for simultaneous five-digit 3D force and \gls{EMG} acquisition with standardized calibration, digit-specific coordinate transformation, and flexible feedback paradigms. 
It enables reproducible, multi-directional characterization of hand kinetics and provides a scalable foundation for studying motor control and developing myoelectric interfaces.

\bibliographystyle{IEEEtran}
\bibliography{references}

@software{BiosigDeviceInterface,
  author       = {Braun, Dominik and Simpetru, Raul},
  title        = {Biosignal-Device-Interface},
  year         = {2024},
  url          = {https://github.com/NsquaredLab/Biosignal-Device-Interface},
  version      = {b2c4dd7cfa2403f30095b33b68ed9ad59fc4f5ff},
}

@article{Pataky2007AHand,
    title = {{A Device for Testing the Intrinsic Muscles of the Hand}},
    year = {2007},
    journal = {Journal of Hand Therapy},
    author = {Pataky, Todd C. and Savescu, Adriana V. and Latash, Mark L. and Zatsiorsky, Vladimir M.},
    number = {4},
    month = {10},
    pages = {345--350},
    volume = {20},
    url = {https://linkinghub.elsevier.com/retrieve/pii/S0894113007000610},
    doi = {10.1197/j.jht.2007.05.002},
    issn = {08941130}
}

@article{Malesevic2019InstrumentedContractions,
    title = {{Instrumented platform for assessment of isometric hand muscles contractions}},
    year = {2019},
    journal = {Measurement Science and Technology},
    author = {Male{\v{s}}evi{\'{c}}, Nebojša and Andersson, Gert and Bj{\"{o}}rkman, Anders and Controzzi, Marco and Cipriani, Christian and Antfolk, Christian},
    number = {6},
    month = {4},
    volume = {30},
    publisher = {Institute of Physics Publishing},
    doi = {10.1088/1361-6501/ab0eae},
    issn = {13616501}
}

@article{Reilly2003IncompleteStudy,
    title = {{Incomplete Functional Subdivision of the Human Multitendoned Finger Muscle Flexor Digitorum Profundus: An Electromyographic Study}},
    year = {2003},
    journal = {Journal of Neurophysiology},
    author = {Reilly, Karen T. and Schieber, Marc H.},
    number = {4},
    month = {10},
    pages = {2560--2570},
    volume = {90},
    url = {https://www.physiology.org/doi/10.1152/jn.00287.2003},
    doi = {10.1152/jn.00287.2003},
    issn = {0022-3077}
}

@article{Kapur2010FingerTask,
    title = {{Finger interaction in a three-dimensional pressing task}},
    year = {2010},
    journal = {Experimental Brain Research},
    author = {Kapur, Shweta and Friedman, Jason and Zatsiorsky, Vladimir M. and Latash, Mark L.},
    number = {1},
    month = {5},
    pages = {101--118},
    volume = {203},
    url = {http://link.springer.com/10.1007/s00221-010-2213-7},
    doi = {10.1007/s00221-010-2213-7},
    issn = {0014-4819}
}

@article{Irwin2013DevelopmentForce,
    title = {{Development and application of a multi-axis dynamometer for measuring grip force}},
    year = {2013},
    journal = {Ergonomics},
    author = {Irwin, C.B. and Towles, J.D. and Radwin, R.G.},
    number = {12},
    month = {12},
    pages = {1841--1849},
    volume = {56},
    url = {http://www.tandfonline.com/doi/abs/10.1080/00140139.2013.847212},
    doi = {10.1080/00140139.2013.847212},
    issn = {0014-0139}
}

@inproceedings{Castellini2014AForces,
    title = {{A wearable low-cost device based upon Force-Sensing Resistors to detect single-finger forces}},
    year = {2014},
    booktitle = {5th IEEE RAS/EMBS International Conference on Biomedical Robotics and Biomechatronics},
    author = {Castellini, Claudio and Ravindra, Vikram},
    month = {8},
    pages = {199--203},
    publisher = {IEEE},
    url = {https://ieeexplore.ieee.org/document/6913776},
    isbn = {978-1-4799-3128-6},
    doi = {10.1109/BIOROB.2014.6913776}
}

@article{Ross2023DesignMeasurement,
    title = {{Design of a Custom Force Fixture Fitting Miniature Load Cells for Individual Finger Force Measurement}},
    year = {2023},
    journal = {Proceedings of the Human Factors and Ergonomics Society Annual Meeting},
    author = {Ross, Lauren and Kutchey, Nichole and Rayess, Nassif and Conrad, Megan O.},
    number = {1},
    month = {9},
    pages = {379--384},
    volume = {67},
    publisher = {SAGE Publications Inc.},
    url = {https://journals.sagepub.com/doi/10.1177/21695067231192434},
    doi = {10.1177/21695067231192434},
    issn = {1071-1813}
}

@article{Grison2024IntramuscularControl,
    title = {{Intramuscular High-Density Micro-Electrode Arrays Enable High-Precision Decoding and Mapping of Spinal Motor Neurons to Reveal Hand Control}},
    year = {2024},
    author = {Grison, Agnese and Pereda, Jaime Ibanez and Muceli, Silvia and Kundu, Aritra and Baracat, Farah and Indiveri, Giacomo and Donati, Elisa and Farina, Dario},
    month = {10},
    url = {http://arxiv.org/abs/2410.11016},
    arxivId = {2410.11016}
}

@article{DelVecchio2020,
    title = {{Tutorial: Analysis of motor unit discharge characteristics from high-density surface EMG signals}},
    year = {2020},
    journal = {Journal of Electromyography and Kinesiology},
    author = {Del Vecchio, A. and Holobar, A. and Falla, D. and Felici, F. and Enoka, R.M. and Farina, D.},
    month = {8},
    pages = {102426},
    volume = {53},
    publisher = {Elsevier},
    url = {https://doi.org/10.1016/j.jelekin.2020.102426 https://linkinghub.elsevier.com/retrieve/pii/S1050641120300419},
    doi = {10.1016/j.jelekin.2020.102426},
    issn = {10506411},
    pmid = {32438235}
}

@article{Zajac1992,
    title = {{How musculotendon architecture and joint geometry affect the capacity of muscles to move and exert force on objects: A review with application to arm and forearm tendon transfer design}},
    year = {1992},
    journal = {The Journal of Hand Surgery},
    author = {Zajac, Felix E.},
    number = {5},
    month = {9},
    pages = {799--804},
    volume = {17},
    url = {https://linkinghub.elsevier.com/retrieve/pii/036350239290445U},
    doi = {10.1016/0363-5023(92)90445-U},
    issn = {03635023}
}

@incollection{Hepp-Reymond1996PrecisionHumans,
    title = {{Precision Grip in Humans}},
    year = {1996},
    booktitle = {Hand and Brain},
    author = {Hepp-Reymond, Marie-Claude and Huesler, Erhard J. and Maier, Marc A.},
    pages = {37--68},
    publisher = {Elsevier},
    url = {https://linkinghub.elsevier.com/retrieve/pii/B9780127594408500062},
    doi = {10.1016/B978-012759440-8/50006-2}
}

@article{simpetru_identification_2024,
    title = {Identification of {Spared} and {Proportionally} {Controllable} {Hand} {Motor} {Dimensions} in {Motor} {Complete} {Spinal} {Cord} {Injuries} {Using} {Latent} {Manifold} {Analysis}},
    volume = {32},
    copyright = {https://creativecommons.org/licenses/by/4.0/legalcode},
    issn = {1534-4320, 1558-0210},
    url = {https://ieeexplore.ieee.org/document/10703163/},
    doi = {10.1109/TNSRE.2024.3472063},
    language = {en},
    urldate = {2026-04-20},
    journal = {IEEE Transactions on Neural Systems and Rehabilitation Engineering},
    author = {Sîmpetru, Raul C. and Souza De Oliveira, Daniela and Ponfick, Matthias and Del Vecchio, Alessandro},
    year = {2024},
    pages = {3741--3750},
}

@article{simpetru_proportional_2023,
    title = {Proportional and {Simultaneous} {Real}-{Time} {Control} of the {Full} {Human} {Hand} {From} {High}-{Density} {Electromyography}},
    volume = {31},
    issn = {1534-4320},
    url = {https://ieeexplore.ieee.org/document/10182328/},
    doi = {10.1109/TNSRE.2023.3295060},
    journal = {IEEE Transactions on Neural Systems and Rehabilitation Engineering},
    publisher = {Institute of Electrical and Electronics Engineers Inc.},
    author = {Sîmpetru, Raul C. and März, Michael and Del Vecchio, Alessandro},
    year = {2023},
    pages = {3118--3131},
}

@article{nestler_reliability_2019,
    title = {Reliability and validity of the finger flexor dynamometer},
    volume = {24},
    issn = {1758-9983},
    url = {https://journals.sagepub.com/doi/10.1177/1758998319859382},
    doi = {10.1177/1758998319859382},
    number = {3},
    journal = {Hand Therapy},
    publisher = {SAGE Publications Ltd},
    author = {Nestler, Kai and Rohde, Ulrich and Becker, Benjamin and Waldeck, Stephan and Veit, Daniel A and Leyk, Dieter},
    month = sep,
    year = {2019},
    pages = {82--90},
}

@article{li_multiaxis_2025,
    title = {Multiaxis {Force}/{Torque} {Sensor} {Technologies}: {Design} {Principles} and {Robotic} {Force} {Control} {Applications}: {A} {Review}},
    volume = {25},
    issn = {1558-1748},
    shorttitle = {Multiaxis {Force}/{Torque} {Sensor} {Technologies}},
    url = {https://ieeexplore.ieee.org/document/10755023},
    doi = {10.1109/JSEN.2024.3495507},
    number = {3},
    urldate = {2026-04-28},
    journal = {IEEE Sensors Journal},
    author = {Li, Shuhan and Xu, Jinli},
    month = feb,
    year = {2025},
    pages = {4055--4069},
}

@article{kebede_decoupled_2019,
    address = {Basel, Switzerland},
    title = {Decoupled {Six}-{Axis} {Force}-{Moment} {Sensor} with a {Novel} {Strain} {Gauge} {Arrangement} and {Error} {Reduction} {Techniques}},
    volume = {19},
    issn = {1424-8220},
    doi = {10.3390/s19133012},
    language = {eng},
    number = {13},
    journal = {Sensors},
    author = {Kebede, Getnet Ayele and Ahmad, Anton Royanto and Lee, Shao-Chun and Lin, Chyi-Yeu},
    month = jul,
    year = {2019},
    pages = {3012},
}

@article{al-mai_design_2018,
    title = {Design, {Development} and {Calibration} of a {Lightweight}, {Compliant} {Six}-{Axis} {Optical} {Force}/{Torque} {Sensor}},
    volume = {18},
    issn = {1558-1748},
    url = {https://ieeexplore.ieee.org/abstract/document/8410875},
    doi = {10.1109/JSEN.2018.2856098},
    number = {17},
    urldate = {2026-04-28},
    journal = {IEEE Sensors Journal},
    author = {Al-Mai, Osama and Ahmadi, Mojtaba and Albert, Jacques},
    month = sep,
    year = {2018},
    pages = {7005--7014},
}

\end{document}